\documentclass[12pt]{article}

\usepackage{amsfonts,amsmath,amssymb,accents,mathrsfs}
\usepackage[retainorgcmds]{IEEEtrantools}
\usepackage{multirow}
\usepackage{multicol}
\usepackage{tikz}
\usepackage{graphicx,color}
\usepackage{booktabs}
\usepackage{placeins}
\usepackage[colorlinks,linkcolor=Blue,citecolor=Blue,urlcolor=Blue,bookmarks,
bookmarksnumbered]{hyperref}
\usepackage[nosort]{cite}
\usepackage[scaled=0.85]{helvet}
\usepackage[T1]{fontenc}
\usepackage[utf8]{inputenc}
\usepackage{cancel}

\definecolor{Red}    {rgb}{0.90,0.00,0.12} 
\definecolor{Blue}   {rgb}{0.00,0.00,1.00} 
\definecolor{Green}  {rgb}{0.10,0.70,0.10} 
\definecolor{Turque} {rgb}{0.00,0.65,0.85} 
\definecolor{Orange} {rgb}{1.00,0.50,0.15} 
\definecolor{Magenta}{rgb}{1.00,0.00,1.00} 
\definecolor{Gold}   {rgb}{1.00,0.75,0.25} 
\definecolor{Seaweed}{rgb}{0.01,0.24,0.09} 
\definecolor{Purple} {rgb}{0.50,0.25,0.55} 
\definecolor{Brown}  {rgb}{0.43,0.26,0.32} 
\definecolor{grey1}  {rgb}{0.20,0.20,0.20} 
\definecolor{grey2}  {rgb}{0.40,0.40,0.40} 
\definecolor{grey3}  {rgb}{0.60,0.60,0.60} 
\definecolor{grey4}  {rgb}{0.80,0.80,0.80} 
\definecolor{grey5}  {rgb}{0.90,0.90,0.90} 

\def\a{{\alpha}}
\def\b{{\beta}}
\def\g{{\gamma}}
\def\d{{\delta}}

\def\th{{\theta}}

\def\ad{{\dot{\alpha}}}
\def\bd{{\dot{\beta}}}

\def\thd{{\bar{\theta}}}

\def\N{{\mathcal{N}}}

\def\J{{\mathcal{J}}}
\def\T{{\mathcal{T}}}

\def\W{{\mathcal{W}}}

\def\Ysf{{\textsf{Y}}}

\def\D{{\rm D}}
\def\Dd{{\bar{\rm D}}}
\def\pa{\partial}

\def\ff#1-#2{\frac{#1}{#2}}
\def\tff#1-#2{\tfrac{#1}{#2}}

\def\be{\begin{equation}}
\def\ee{\end{equation}}
\def\bea{\begin{IEEEeqnarray*}}
\def\eea{\end{IEEEeqnarray*}}
\def\n{\IEEEyesnumber}
\def\sn{\IEEEyessubnumber}

\makeatletter
\def\section{\@startsection{section}{1}{\z@}
              {3ex plus-1ex minus-.2ex}{1pt plus1pt}
              {\large\sf\bfseries\boldmath}}
\def\subsection{\@startsection{subsection}{2}{\z@}
              {1.5ex plus-1ex minus-.2ex}{0.01pt plus1pt}{\sf\slshape}}
\def\subsubsection{\@startsection{subsubsection}{3}{\z@}
              {1.5ex plus-1ex minus-.2ex}{0.01pt plus0.2pt}{\sf\boldmath}}
\def\paragraph{\@startsection{paragraph}{4}{\z@}
              {.75ex \@plus.5ex \@minus.2ex}{-2mm}{\sf\bfseries\boldmath}}
\makeatother

   \parskip\medskipamount     \lineskip=0pt
\thicklines
\begin{document}
\thispagestyle{empty}
\noindent{\small
\vspace*{6mm}
\begin{center}
{\large \bf 
Progress on cubic interactions of arbitrary superspin supermultiplets
via gauge invariant supercurrents
\vspace{3ex}
} \\   [9mm] {\large { 
S.\ James Gates Jr.\footnote{sylvester\_gates@brown.edu}$^{a,b}$
and K.\ Koutrolikos\footnote{konstantinos\_koutrolikos@brown.edu}$^{b}$ }
}
\\*[8mm]
\emph{
\centering
$^{a}$Brown Theoretical Physics Center
\\[6pt]
$^{b}$Department of Physics, Brown University,
\\[1pt]
Box 1843, 182 Hope Street, Barus \& Holley 545,
Providence, RI 02912, USA
}
 $$~~$$
  $$~~$$
 \\*[-8mm]
{ ABSTRACT}\\[4mm]
\parbox{142mm}{\parindent=2pc\indent\baselineskip=14pt plus1pt
We consider cubic interactions of the form $s-\Ysf-\Ysf$ between a massless integer superspin
$s$
supermultiplet and two massless arbitrary integer or half integer superspin $\Ysf$
supermultiplets. We focus on non-minimal interactions generated by
gauge invariant supercurrent multiplets which are bilinear in the superfield strength
of the superspin $\Ysf$ supermultiplet. We find two types of consistent supercurrents.
The first one corresponds to conformal integer superspin $s$ supermultiplets, exist only
for even values of $s, s=2\ell+2$, for arbitrary values of $\Ysf$ and it is unique. The second one,
corresponds to Poincar\'e integer superspin $s$ supermultiplets, exist for arbitrary values of $s$
and $\Ysf$.
}
\end{center}
$$~~$$
\vfill
\vfill
\clearpage
%
\section{Introduction}
\label{intro}
The theory of massless higher spin fields can be understood as an attempt to
understand and classify the potential list of symmetries emerging from string
theory at high enough energy scales. As such, it is only natural to enhance higher spin
symmetry with supersymmetry which is another key ingredient of strings. This line of thought
inevitably leads to the study of higher spin irreducible representations of the super-Poincar\'{e}
group. The free theory of massless, higher \emph{superspins} for flat spacetime, its AdS and conformal versions
has been developed \cite{Curtright:1979uz,Vasiliev:1980as,Kuzenko:1993jp,Kuzenko:1993jq,Kuzenko:1994dm,
Gates:1996my,Gates:1996xs,Gates:2013rka,Gates:2017hmb}\footnote{Recently, a manifestly
supersymmetric description of continuous spin representations has been proposed \cite{Buchbinder:2019esz}
but for the purpose of this paper we will not include such representations under the label of higher spins.}.

The problem of finding consistent interactions involving these higher spin supermultiplets is non-trivial
as in the case of non-supersymmetric higher spin theories. At present, a wide class of cubic
interactions of the type $\Ysf-0-0$,  between arbitrary integer or half-integer superspin (\Ysf) supermultiplets and 
various matter supermultiplets ($\Ysf=0$) is known \cite{Kuzenko:2017ujh,Buchbinder:2017nuc,
Hutomo:2017phh,Hutomo:2017nce,Koutrolikos:2017qkx,Buchbinder:2018wwg,
Buchbinder:2018nkp,Hutomo:2018tjh,Buchbinder:2018gle}. A more general class of non-minimal cubic
interactions of the type $(s+\tff1-2)-\Ysf-\Ysf$ has been discovered in \cite{Buchbinder:2018wzq}.
These are interactions between a half-integer superspin $\Ysf=s+\tff1-2$ supermultiplet and two
arbitrary superspin $\Ysf$ (integer of half-integer) supermultiplets and are gauge invariant and 
the corresponding higher spin supercurrent which generates the interaction is quadratic in the superfield strengths
of the superspin $\Ysf$ supermultiplets. It was also found that such interactions
exist for all values of $s$,
but superspin $\Ysf$ (can be integer or half-integer) is bounded by $\ff s-2$ [$\Ysf\leq\ff s-2$].
This was understood as a supersymmetric higher spin generalization of the Weinberg-Witten theorem.

For non-supersymmetric theories it has been demonstrated in \cite{Berends:1985xx,Gelfond:2006be}
that such gauge invariant higher spin interactions, despite their non-minimal nature
are interesting because if they exist they are unique. 
One may think that their results could be easily supersymmetrized and possibly with an arbitrary number of supersymmetries.
However, that would be true only for on-shell supersymmetries. That means the invariance of the
interacting action under supersymmetry will not be automatic but has to be checked and it will hold only if we use the
equations of motion (on-shell). Furthermore, the supersymmetry transformations that one can write between
the bosonic and fermionic higher spin fields will not be consistent with the
supersymmetry algebra unless we use the equations of motion. Therefore, if one demands off-shell supersymmetry,
in other words requires supersymmetry to be manifest, then one is forced to use the entire supermultiplet.
The supermultiplet includes not only the higher spin fields that describe the dynamics, but also an additional set
of auxiliary fields. These auxiliary fields have no dynamics and do not describe any on-shell propagating degrees of freedom.
Their purpose is to balance the off-shell bosonic degrees of freedom with the off-shell
fermionic degrees of freedom.
Additionally, they make the supersymmetry transformation
laws of all the fields (dynamical and auxiliary) satisfy the supersymmetry algebra without the need of the equations
of motion and the invariance of the action under supersymmetry is manifest. All these conditions are met automatically if one
adopts the superspace description of the theory using superfields.
Increasing the number of off-shell supersymmetries, we have to increase the number of these auxiliary
fields in an non-trivial way thus changing the structure of superspace that underlies the theory.

In \cite{Buchbinder:2018wzq} for $4D,~\N=1$ we managed to find the appropriate structure to describe
manifestly supersymmetric, gauge invariant, interactions between a half integer superspin supermultiplet ($\Ysf=s+1/2$)\footnote{The half 
integer superspin supermultiplet describes spins $j=s+1$ and $j=s$.} with two supermultiplets of arbitrary superspin $\Ysf$\footnote{The
supermultiplet with arbitrary (integer or half-integer) superspin $\Ysf$ describes spins $j=\Ysf+1/2$ and $j=\Ysf$.}.
In this letter, we continue the investigation of non-minimal cubic interactions among massless higher spin
supermultiplets
and prove the existence of a new wide class of consistent interactions of the type $s-\Ysf-\Ysf$
between an integer superspin ($\Ysf=s$) supermultiplet and two arbitrary (integer or half integer) superspin
$\Ysf$ supermultiplets. We find two types of such interactions which are related with the conformal or
Poincar\'e nature of the integer superspin $s$ supermultiplet.
For the first one, the integer superspin supercurrent satisfies conservation equations that correspond to
a conformal integer superspin $\Ysf=s$ supermultiplet, it is unique and exist only for even values
of $s$ ($s=2\ell+2$). For the second one, the integer superspin supercurrent
satisfies conservation equations that correspond to a Poincar\'e integer superspin $s$ supermultiplet, it is not unique
and there is no selection rule, it exist for all values of $s$. Most importantly and in a big contrast with the results of
\cite{Buchbinder:2018wzq}, there is no constraint on the values of $\Ysf$. These interactions exist for arbitrary $\Ysf$.

This letter is organized as follows. In section 2, we review basic elements of massless,
higher spin supermultiplets. We briefly provide the list of superfields that participate in the description
of various conformal or Poincar\'e, integer and half-integer superspin supermultiplets and their gauge transformations.
The consistency of cubic interactions with the free theory gauge transformations will generate the various conservation laws
for the higher spin supercurrent multiplets. Sections 3 and 4 present the construction of 
higher spin, gauge invariant, supercurrents which generate consistent (respect the previous conservation laws)
$s-\Ysf-\Ysf$ cubic interactions,
for the conformal and Poincar\'e cases. Finally, section 5 presents a summary of our results.
%
\section{Higher spin supermultiplets, superfield strengths and conservation equations}
\label{rceq}
We consider cubic interactions of irreducible, $4D,~\N=1$, higher spin supermultiplets. The
off-shell, superspace, description of their free theory was given first in \cite{Kuzenko:1993jp,Kuzenko:1993jq} by proposing a set
of various superfields, including constrained ones, and their gauge transformations. Based on this proposition,
the action principle was uniquely determined and led to the correct on-shell equations of motion for the
various field strength supertensors.
It was also commented that the various constrained superfields could be expressed in terms of unconstrained prepotentials with appropriate
gauge transformations, thus solving the constraints.
Build upon these foundational results,
in \cite{Gates:2013rka} later it was shown an alternative path
exists. A careful consideration of the massless limit of massive higher spin supermultiplets
will lead to a description of massless higher spin supermultiplets
in terms of the unconstrained superfields and correctly generate their gauge transformations.
Furthermore, a detailed analysis
of the component structure of the theory was given. This includes the field spectrum of the theory,
the component action and the set of supersymmetry transformations for all components, which leave the action invariant.
For the purpose of our discussion we review\footnote{We follow \cite{Gates:2013rka} and we use the
conventions of ``\emph{Superspace}'' \cite{Gates:1983nr}.} the basic results:
\begin{enumerate}
\item The integer superspin $\Ysf=s$ ($s\geq1$) supermultiplets $(s+1/2 ,
s)$\footnote{On-shell they
describe the propagation of helicities $\pm (s+1/2)$ and
$\pm s$.}
are described by a pair of superfields
$\Psi_{\a(s)\ad(s-1)}$\footnote{The notation $\a(k)$ is a shorthand for k
undotted symmetric indices $\a_1\a_2\dots\a_k$.
Similarly for dotted indices.}
 and $V_{\a(s-1)\ad(s-1)}$ (real) with the following lowest
order gauge transformations
\bea{l}\n\label{hstr1}
\d_0\Psi_{\a(s)\ad(s-1)}=-\D^2L_{\a(s)\ad(s-1)}+\tfrac{1}{(s-1)!}\Dd_{(\ad_{
s-1}}\Lambda_{\a(s)\ad(s-2))}~,\sn\label{PsiP}\\
\d_0 V_{\a(s-1)\ad(s-1)}=\D^{\a_s}L_{\a(s)\ad(s-1)}+\Dd^{\ad_s}\bar{L}_{\a(s-1)\ad(s)}\sn~.
\eea
Off-shell, this supermultiplet carries $8s^2+8s+4$ bosonic and equal number of
fermionic degrees for freedom.
\item The half-integer superspin $\Ysf=s+1/2$ supermultiplets $(s+1 , s+1/2)$
have two descriptions. The first one ($s\geq1$) uses the pair of superfields 
$H_{\a(s)\ad(s)}$ (real) and $\chi_{\a(s)\ad(s-1)}$ with the following lowest order gauge
transformations
\bea{l}\n\label{hstr2}
\d_0
H_{\a(s)\ad(s)}=\tfrac{1}{s!}\D_{(\a_s}\bar{L}_{\a(s-1))\ad(s)}-\tfrac{1}{s!}
\Dd_{(\ad_s}L_{\a(s)\ad(s-1))}\sn\label{HP1}~,\\
\d_0\chi_{\a(s)\ad(s-1)}=\Dd^2L_{\a(s)\ad(s-1)}+\D^{\a_{s+1}}\Lambda_{\a(
s+1)\ad(s-1)}~.\sn
\eea
This supermultiplet, off-shell describes $8s^2+8s+4$ bosonic and equal number fermions.
The second formulation ($s\geq2$) has the same $H_{\a(s)\ad(s)}$ as previously but
a different compensating superfield
$\chi_{\a(s-1)\ad(s-2)}$ with gauge transformations
\bea{l}\n\label{hstr3}
\d_0
H_{\a(s)\ad(s)}=\tfrac{1}{s!}\D_{(\a_s}\bar{L}_{\a(s-1))\ad(s)}-\tfrac{1}{s!}
\Dd_{(\ad_s}L_{\a(s)\ad(s-1))}\sn\label{HP2}~,\vspace{1ex}\\
\d_0\chi_{\a(s-1)\ad(s-2)}=\Dd^{\ad_{s-1}}\D^{\a_s}L_{\a(s)\ad(s-1)}+\tfrac{
s-1}{s}\D^{\a_s}\Dd^{\ad_{s-1}}L_{\a(s)\ad(s-1)}\sn\\
\hspace{17ex}
+\tfrac{1}{(s-2)!}\Dd_{(\ad_{s-2}}J_{\a(s-1)\ad(s-3))}~.
\eea
This supermultiplet carries $8s^2+4$ off-shell bosonic and equal number of
fermionic degrees of freedom.
\end{enumerate}
The free theory actions (quadratic in the superfields and up to two spacetime derivatives) that describe
the above irreducible representations are uniquely determined by the gauge symmetries.
The explicit expression of these free actions in terms of the corresponding unconstrained superfields
can be found in \cite{Gates:2013rka,Gates:2013ska,Koutrolikos:2015lqa}. For the case of integer
superspin supermultiplets, which will be our focus in this letter, the free action is (up to an overall constant c):
\bea{ll}
S=\int d^8z&\left\{-\frac{1}{2}c\Psi^{\a(s)\ad(s-1)}\Dd^2
\Psi_{\a(s)\ad(s-1)}+c.c.\right.\\
&~~+c\Psi^{\a(s)\ad(s-1)}\Dd^{\ad_s}\D_{\a_s}\bar{\Psi
}_{\a(s-1)\ad(s)}\\
&~~-c V^{\a(s-1)\ad(s-1)}\D^{\a_s}\Dd^2\Psi_{\a(s)\ad(s-1
)}+c.c.\IEEEyesnumber\\
&~~\left.+\frac{1}{2}c V^{\a(s-1)\ad(s-1)}\D^{\g}\Dd^2\D_{
\g}V_{\a(s-1)\ad(s-1)}\right\}~.
\eea

The physical and propagating degrees of freedom for
massless integer and half-integer superspins are described by
superfield strengths $\W_{\a(2s)}$ and $\W_{\a(2s+1)}$
respectively. They are defined in the following way:
\bea{ll}\n\label{sfsW}
\Ysf=s+1/2 : ~&~\W_{\a(2s+1)}\sim\Dd^2\D_{(\a_{2s+1}}\pa_{\a_{2s}}{}^{\ad_s}
\pa_{\a_{2s-1}}{}^{\ad_{s-1}}\dots\pa_{\a_{s+1}}{}^{\ad_{1}}H_{\a(s))\ad(s)}~,\sn
\vspace{2ex}\\
\Ysf=s : &~\W_{\a(2s)}\sim\Dd^2\D_{(\a_{2s}}\pa_{\a_{2s-1}}{}^{\ad_{s-1}}
\pa_{\a_{2s-2}}{}^{\ad_{s-2}}\dots\pa_{\a_{s+1}}{}^{\ad_{1}}\Psi_{\a(s))\ad(s-1)}
\sn
\eea
and they are invariant with respect the respective gauge symmetries mentioned above.
Their characteristic feature is to have a special index structure, \emph{i.e.}
they have only one type of index and $2\Ysf$ of them. Moreover, they are chiral
\bea{l}
\Dd_{\bd}\W_{\a(2s+1)}=0~~~,~~~~\Dd_{\bd}\W_{\a(2s)}=0\n\label{cW}
\eea
and on-shell they satisfy the following equations of motion
\bea{l}
\D^{\b}\W_{\b\a(2s)}=0~~~,~~~\D^{\b}\W_{\b\a(2s-1)}=0~.\n\label{Weom}
\eea
At the component level they include the bosonic and fermionic higher spin field
strengths.

Notice that in all cases, we need two superfields to describe the corresponding
higher superspin supermultiplet. The first one is associated with the superfield strength
and plays the role of the (pre)potential and the second one (compensator) is required in order to
write a two derivative, manifestly super-Poincar\'e, invariant action. 
However, one can consider conformal higher superspin supermultiplets. The Lagrangian description of such
representations is given purely in terms of the superfield strengths, as described above, so it includes higher
derivatives. The free action is:
\bea{l}
S=\int d^6z ~\W^{\a(2\Ysf)}\W_{\a(2\Ysf)}~+c.c.\n
\eea
For these theories we require only one superfield, the (pre)potential which must be
a primary superfield with appropriate weights\footnote{A quick review of primary superfields can be found in
\cite{Kuzenko:2017ujh}.}. Its gauge 
transformation is determined by the largest symmetry that preserves the superfield strength:
\begin{enumerate}
\item The conformal integer superspin $\Ysf=s$, being described by superfield $\Psi_{\a(s)\ad(s-1)}$,
which is primary with conformal weights $(-\ff s-2,-\ff{s-1}-2)$ and has a gauge transformation
\bea{l}\n\label{PsiC}
\d_0\Psi_{\a(s)\ad(s-1)}=\tfrac{1}{s!}~\D_{(\a_s}\Xi_{\a(s-1))\ad(s-1)}
+\tfrac{1}{(s-1)!}~\Dd_{(\ad_{s-1}}\Lambda_{\a(s)\ad(s-2))}~.
\eea
\item The conformal half-integer superspin $\Ysf=s+\ff1-2$, being described by a real primary superfield
$H_{\a(s)\ad(s)}$ with conformal weights $(-\ff s-2, -\ff s-2)$ and gauge transformation
\bea{l}\n\label{HC}
\d_0
H_{\a(s)\ad(s)}=\tfrac{1}{s!}\D_{(\a_s}\bar{L}_{\a(s-1))\ad(s)}-\tfrac{1}{s!}
\Dd_{(\ad_s}L_{\a(s)\ad(s-1))}~.
\eea
\end{enumerate}
Notice that the gauge transformation \eqref{HC} is identical to the corresponding gauge transformations
(\ref{HP1},~\ref{HP2}) of the super-Poincar\'e half-integer superspin representations. However, for the
integer superspin case, the gauge transformation \eqref{PsiC} of the conformal representation
is larger than the corresponding Poincar\'e one \eqref{PsiP}. This difference will be the reason why
for integer superspin interactions we find two sets of conserved supercurrents whereas for the half-integer
case only one.

To make this clear, let's consider cubic interactions of type $\Ysf-\Ysf_1-\Ysf_2$
between supermultiplets with superspin values $\Ysf,~\Ysf_1,~\Ysf_2$.
Assuming that such interactions exist, they are local and manifestly super-Poincar\'e or super-conformal then they can be written
in the following form:
\begin{enumerate}
\item For $\Ysf=s$
\bea{l}\n
\text{Poincar\'e}~:~S_{s-\Ysf_1-\Ysf_2}=\displaystyle\int d^8z\left\{[\Psi^{\a(s)\ad(s-1)}\J_{\a(s)\ad(s-1)}+c.c.]
+V^{\a(s-1)\ad(s-1)}\T_{\a(s-1)\ad(s-1)}\right\}\hspace{7ex}\sn\label{ciP}\vspace{2ex}\\
\text{conformal}~:~S_{s-\Ysf_1-\Ysf_2}=\displaystyle\int d^8z~\Psi^{\a(s)\ad(s-1)}\J_{\a(s)\ad(s-1)}+c.c.\sn\label{ciC}
\eea
where the higher spin supercurrent $\J_{\a(s)\ad(s-1)}$ and the real higher spin supertrace $\T_{\a(s-1)\ad(s-1)}$
are bilinear in the $(\Ysf_1,\Ysf_2)$ supermultiplets. In order for the added cubic interaction
to preserve the degrees of freedom of the free theory and not introduce new ones, it must respect
the free theory gauge transformations  \eqref{hstr1} and \eqref{PsiC}. Therefore, the supercurrent multiplet
must satisfy the following conservation equations:
\bea{l}\n
\text{Poincar\'e~:~}\D^2\J_{\a(s)\ad(s-1)}=\tfrac{1}{s!}~\D_{(\a_s}\T_{\a(s-1))\ad(s-1)}~~,~~
\Dd^{\ad_{s-1}}\J_{\a(s)\ad(s-1)}=0\sn\label{ceP}\vspace{2ex}\\
\text{conformal~:~}\D^{\a_{s}}\J_{\a(s)\ad(s-1)}=0
~~,~~\Dd^{\a_{s-1}}\J_{\a(s)\ad(s-1)}=0~~.\sn\label{ceC}
\eea
For the conformal case, the supercurrent has an extra condition. It must be primary with weights $(1+\ff s-2,1+\ff{s-1}-2)$.
\item For $\Ysf=s+\ff1-2$
\bea{l}\n\label{cihi}
\text{Poincar\'e~I}~:~S_{(s+\tff1-2)-\Ysf_1-\Ysf_2}=\displaystyle\int d^8z\left\{H^{\a(s)\ad(s)}\J_{\a(s)\ad(s)}+
[\chi^{\a(s)\ad(s-1)}\T_{\a(s)\ad(s-1)}+c.c.]\right\}\hspace{7ex}\sn\vspace{2ex}\\
\text{Poincar\'e~II}~:~S_{(s+\tff1-2)-\Ysf_1-\Ysf_2}=\displaystyle\int d^8z\left\{H^{\a(s)\ad(s)}\J_{\a(s)\ad(s)}+
[\chi^{\a(s-1)\ad(s-2)}\T_{\a(s-1)\ad(s-2)}+c.c.]\right\}\hspace{7ex}\sn\vspace{2ex}\\
\text{conformal}~:~S_{(s+\tff1-2)-\Ysf_1-\Ysf_2}=\displaystyle\int d^8z ~H^{\a(s)\ad(s)}\J_{\a(s)\ad(s)}\sn
\eea
where the real higher spin supercurrent and the higher spin supertrace
satisfy the following conservation equations
\bea{l}\n\label{cehi}
\text{Poincar\'e~I~:~}\Dd^{\ad_s}\J_{\a(s)\ad(s)}=\Dd^2\T_{\a(s)\ad(s-1)}~~,~~\D_{(\a_{s+1}}\T_{\a(s))\ad(s-1)}=0\sn\label{cePI}\vspace{2ex}\\
\text{Poincar\'e~II~:~}\Dd^{\ad_s}\J_{\a(s)\ad(s)}=-\tff{1}-{s!(s-1)!}\D_{(\a_s}\Dd_{(\ad_{s-1}}\T_{\a(s-1))\ad(s-2))}
-\tff{s-1}-{s!s!}\Dd_{(\ad_{s-1}}\D_{(\a_s}\T_{\a(s-1))\ad(s-2))},
~~~~~~\sn\label{cePII}\\
\hspace{13.5ex}\Dd^{\ad_{s-2}}\T_{\a(s-1))\ad(s-2)}=0\vspace{2ex}\\
\text{conformal~:~}\D^{\a_{s}}\J_{\a(s)\ad(s-1)}=0~.\sn\label{ceChi}
\eea
Additionally for the conformal case the supercurrent must be primary with weights $(1+\ff s-2,1+\ff{s}-2)$.
\end{enumerate}
One has to keep in mind that the supercurrent and supertrace pair which generates the cubic interaction, in general
is not unique. One can consider improvement terms and produce an infinite family
of equivalent $\{\J,\T\}$ pairs. For example, using this freedom one can exchange conservation equations \eqref{cePI} and \eqref{cePII} 
\cite{Buchbinder:2018wwg} and reveal the duality that exists between the two super-Poincar\'e half-integer superspin supermultiplets.
In other cases, it is possible to use the improvement terms in order to make the supertrace vanish ($\T=0$).
For these cases, there is no distinction between the Poincar\'e and conformal supercurrents if $\Ysf=s+\ff1-2$
at the level of cubic interactions \eqref{cihi} and conservation equations \eqref{cehi}.
Of course, one also has to check the primary nature of the  \emph{minimal}
\footnote{This is the new supercurrent acquired by the addition of the 
improvement terms that make the supertrace vanish: $\{\J,\T\}\sim\{\J_{\text{minimal}},0\}$} supercurrent.
However, using arguments similar to \cite{Craigie:1983fb}, one may connect the proper transformations under conformal symmetry with the
conformal conservation equations \eqref{ceChi}. On the other hand for 
$\Ysf=s$, one can still distinguish between the Poincar\'e and conformal supercurrents since the left-hand sides of
conservation equations \eqref{ceP} and \eqref{ceC} are different.

In previous works \cite{Kuzenko:2017ujh,Buchbinder:2017nuc,
Hutomo:2017phh,Hutomo:2017nce,Koutrolikos:2017qkx,Buchbinder:2018wwg,
Buchbinder:2018nkp,Buchbinder:2018gle,Hutomo:2018tjh} a variety of cubic interactions between
arbitrary, massless, integer or half-integer superspin supermultiplets and massless or massive matter supermultiplets $[s+\ff1-2-0-0~,~s-0-0]$
have been found either by solving the corresponding conservation equations
or using Noether's method with appropriate transformations in order to generate consistent supercurrent multiplets.
Another step was made in \cite{Buchbinder:2018wzq} where new cubic interaction between arbitrary massless half-integer superspin 
supermultiplets and massless integer or half-integer superspin $\Ysf$ supermultiplets $[s+\ff1-2 -\Ysf-\Ysf]$ were found. 
These interactions have two characteristic properties. The first one is that the higher spin supercurrent
can be written in terms of the superfield strengths $\W_{\a(2\Ysf)}$ of the two superspin $\Ysf$ supermultiplets,
hence it is a non-minimal class of 
interactions and the supercurrent is gauge invariant. The second one is that these types of interactions do not exist
for arbitrary $\Ysf$ but only if $\Ysf\leq\ff{s}-2$. In this work, we investigate similar type of interactions for the integer superspin 
supermultiplet $[s-\Ysf-\Ysf]$. We find that such interactions are possible for both the conformal \eqref{ceC} and Poincar\'e cases
\eqref{ceP} with a vanishing supertrace. A surprising distinction from previous results is that there
is no upper bound in the value of superspin $\Ysf$. However, there is an even values of $s$ selection rule for the conformal case. 
%
%
\section[s-Y-Y]{Conformal integer superspin $s$ with arbitrary superspin $\Ysf$:~~$s-\Ysf-\Ysf$}
\label{cs-Y-Y}
Now let's consider the cubic interaction $s-\Ysf-\Ysf$ between a conformal integer superspin $s$
and two arbitrary, massless superspin $\Ysf$ supermultiplets. The interaction, if it exists and assuming
locality and manifest invariance, must take the form:
\bea{l}
S_{s-\Ysf-\Ysf}=\int d^8z~\Psi^{\a(s)\ad(s-1)}\J_{\a(s)\ad(s-1)}+c.c.\n
\eea
where the higher spin supercurrent must satisfy the conservation equations \eqref{ceC}.
Additionally, the supercurrent must be a composite object, quadratic to the superspin $\Ysf$
supermultiplets. Similarly to \cite{Buchbinder:2018wzq}, we further assume that the supercurrent
is gauge invariant under the gauge transformations of superspin $\Ysf$ and can be written in terms of the superfield strength
$\W_{\a(2Y)}$.
A general ansatz that one can write for the supercurrent is:
\bea{l}
\J_{\a(s)\ad(s-1)}=\sum_{p=0}^{s-1}a_{p}~\pa^{(p)}\D\W^{\g(2\Ysf)}~\pa^{(s-1-p)}\W_{\g(2\Ysf)}\n\label{Jans}
\eea
where for clarity we have suppressed all free $\a$ and $\ad$ indices originating from the
strings of partial spacetime derivatives and the spinorial derivative. Also we have suppressed the
symmetrization of all these indices together with the appropriate symmetrization factors. However, we explicitly indicate
the indices of the two superfield strengths which are contracted to each other and do not contribute to the set of free
indices.
 Using the chiral condition \eqref{cW} it is straightforward to show that
\bea{l}\n\label{DJ}
\D^{\a_s}\J_{\a(s)\ad(s-1)}\approx\D^2\left\{
\sum_{p=0}^{s-1}\left[\hspace{0.5ex}\ff{p+1}-{2s}\hspace{0.5ex}a_{p}+\hspace{0.5ex}\ff{s-p}-{2s}\hspace{0.5ex}a_{s-1-p}\right]~
\pa^{(p)}\W^{\g(2\Ysf)}~\pa^{(s-1-p)}\W_{\g(2\Ysf)}
\right\}
\eea 
where the equality symbol ``$\approx$'' means modulo terms that depend on the equations of motion \eqref{Weom}. When we go
on-shell, as we always do when we calculate conservation equations, this symbol can be replaced with the usual equality symbol.  
The conclusion is that in order for this supercurrent to satisfy the conservation equation $\D^{\a_s}\J_{\a(s)\ad(s-1)}=0$
we must choose the coefficients $a_{p}$ such that
\bea{l}
a_{p}~(p+1)~+~a_{s-1-p}~(s-p)=0~~~,~~~~p=0,1,2,...,s-1~.\n\label{rec1}
\eea
Similarly we can show that
\bea{l}\n\label{DdJ}
\Dd^{\ad_{s-1}}\J_{\a(s)\ad(s-1)}\approx i(-1)^{2\Ysf}\hspace{0.3ex}\Dd^2\left\{
\sum_{p=0}^{s-2}\left[\hspace{0.2ex}\ff{s-1-p}-{2(s-1)}\hspace{0.5ex}a_{p}-\hspace{0.5ex}\ff{p+1}-{2(s-1)}\hspace{0.5ex}a_{s-2-p}\right]
\hspace{0ex}\pa^{(p)}\D\W^{\g(2\Ysf)}\hspace{0.3ex}\pa^{(s-2-p)}\D\W_{\g(2\Ysf)}
\right\}~~~~
\eea
hence in order to satisfy the second conservation equation we must choose the coefficients $a_{p}$ such that
\bea{l}
a_{p}~(s-1-p)~-~a_{s-2-p}~(p+1)=0~~~,~~~~p=0,1,2,...,s-2~.\n\label{rec2}
\eea
The system of recursive relations \eqref{rec1} and \eqref{rec2} can be solved only for even values of $s$. 
For that case the solution is unique
\bea{l}
a_{p}=(-1)^{p}\binom{s-1}{p}\binom{s}{p+1}~~,~~p=0,1,...,s-1~~,~~s=2\ell+2 ~~,~~\ell=0,1,2,...~.\n
\eea

The conclusion is that there is a cubic interaction $s-\Ysf-\Ysf$ between a conformal integer
superspin $s$ and two massless, arbitrary integer or half-integer superspin $\Ysf$ supermultiplets but
only for even values of $s$, $s=2\ell+2$. The supercurrent which generates the cubic interaction is
\bea{l}
\J_{\a(2\ell+2)\ad(2\ell+1)}=\sum_{p=0}^{2\ell+1}(-1)^{p}\binom{2\ell+1}{p}\binom{2\ell+2}{p+1}
~\pa^{(p)}\D\W^{\g(2\Ysf)}~\pa^{(2\ell+1-p)}\W_{\g(2\Ysf)}\n\label{csc}
\eea 
and on-shell it satisfies conservation equations \eqref{ceC}. An interesting observation is that
there is no constraint on the value of $\Ysf$. Another interesting remark is about the $\Ysf\to 0$ limit of \eqref{csc}.
If we set by hand $\Ysf=0$ then $\W$ no longer has the interpretation of the superfield strength of a higher spin 
supermultiplet and expressions \eqref{sfsW} are no longer valid. However, $\W$ remains a chiral superfield and as such describes
a matter supermultiplet. Therefore, by setting $\Ysf=0$ in expression \eqref{csc} we recover precisely
the conformal integer superspin supercurrent of a chiral superfield \cite{Buchbinder:2018nkp,Buchbinder:2018wzq}
which also has the even value selection rule for $s$ and generates the $(2\ell+2)-0-0$ interaction.
%
\section[s-Y-Y]{Poincar\'e integer superspin $s$ with arbitrary superspin $\Ysf$:~~$s-\Ysf-\Ysf$}
\label{Ps-Y-Y}
Now let's consider the possibility of a cubic interaction $s-\Ysf-\Ysf$ between a Poincar\'e integer superspin $s$
and two arbitrary, massless superspin $\Ysf$ supermultiplets. With the same assumptions as previously, 
the interaction, if it exists, must take the form
\bea{l}
S_{s-\Ysf-\Ysf}=\int d^8z~\left\{[\Psi^{\a(s)\ad(s-1)}\J_{\a(s)\ad(s-1)}+c.c.]+V^{\a(s-1)\ad(s-1)}\T_{\a(s-1)\ad(s-1)}\right\}\n
\eea
with the conservation equations \eqref{ceP} for the
higher spin supercurrent and the supertrace. The ansatz for the supercurrent
is the same as \eqref{Jans}. Therefore, due to \eqref{DJ} we immediately find that such a supercurrent
tautologically satisfies $\D^2\J_{\a(s)\ad(s-1)}\approx 0$ for arbitrary values of $a_{p}$. Hence, the supertrace must vanish
\bea{l}
\D^2\J_{\a(s)\ad(s-1)}\approx 0~~,~~\forall~ a_{p}\Rightarrow ~\T_{\a(s-1)\ad(s-1)}=0~.\n
\eea
Lastly, we must check the second conservation equation $\Dd^{\ad_{s-1}}\J_{\a(s)\ad(s-1)}= 0$. Using \eqref{DdJ}
we conclude that coefficients $a_p$ must obey \eqref{rec2}. This is the only constraint that
for coefficients $a\_{p}$ for the Poincar\'e case. This condition is not enough to uniquely fix everything.
For example, notice that these recursion relations do not include
$a_{s-1}$, which remains unconstrained. A general solution of \eqref{rec2} is
\bea{l}
a_{p}=d~\binom{s-1}{p}\binom{s+2\kappa-2}{p+\kappa}^n ~~~,~~~p=0,1,...,s-2\vspace{1ex}\\
a_{s-1}=c
\eea
for arbitrary $c,~d,~\kappa,~n$ and $s$.
Hence, there exist a family of such supercurrents that can generate
the cubic interactions between Poincar\'e integer superspin $s$ and two arbitrary superspin supermultiplets.
They take the following form:
\bea{l}
\J_{\a(s)\ad(s-1)}=c~\pa^{(s-1)}\D\W^{\g(2\Ysf)}~\W_{\g(2\Ysf)}~+~
d~\sum_{p=0}^{s-2}\binom{s-1}{p}\binom{s+2\kappa-2}{p+\kappa}^n~
\pa^{(p)}\D\W^{\g(2\Ysf)}~\pa^{(s-1-p)}\W_{\g(2\Ysf)}~.~~~~~\n\label{JP}
\eea
Unlike the previous result for the conformal case, there is no $s$-selection rule and the supercurrent exist for all values of $s$.
Moreover, this result holds for all values of $\Ysf$, similar to the conformal result. Following the arguments of previous section
we can take the $\Ysf\to 0$ limit in order to recover the integer superspin supercurrent of a chiral supermultiplet.
By setting $\Ysf=0$ and interpreting $\W$ as a chiral superfield $\Phi$, we get precisely the result found in \cite{Buchbinder:2018wzq}
\footnote{In \cite{Buchbinder:2018wzq} only the corresponding to the first term of \eqref{JP} was considered
$\pa^{(s-1)}\D\Phi~\Phi$. That is because the second term would correspond to an improvement term.}.
%
\section{Summary}
\label{outro}
To summarize our results we consider cubic interactions $s-\Ysf-\Ysf$, between one massless integer superspin $s$ supermultiplet
and two massless arbitrary superspin $\Ysf$ supermultiplets. Specifically, we focus on cubic interactions that are generated by gauge
invariant supercurrent multiplets with respect to the gauge symmetry of the two superspin $\Ysf$ supermultiplets. 
For this reason we consider
higher spin supercurrents and supertraces that are composite objects, written in terms of the
superspin $\Ysf$ superfield strength $\W_{\gamma(2\Ysf)}$. A general ansatz for such an integer superspin supercurrent 
$\J_{\a(s)\ad(s-1)}$ can be written \eqref{Jans} and we checked its compatibility with the appropriate conservation equations.
The integer superspin $\Ysf=s$ supermultiplet can be either conformal \eqref{PsiC} or Poincar\'e \eqref{hstr1}
hence the cubic interactions could be of the form \eqref{ciC} or \eqref{ciP} and the supercurrent multiplet must satisfy the conservation 
equations \eqref{ceC} or \eqref{ceP}. For both cases we find a non-trivial supercurrent:
\begin{enumerate}
\item For the conformal case, we find that the integer superspin supercurrent is uniquely fixed \eqref{csc} 
by the conservation equations. Furthermore, the supercurrent and therefore the cubic interaction exist for all values of superspin $\Ysf$
but only for even values of $s,~s=2\ell+2$. Moreover, by setting $\Ysf=0$ we recover the result of a conformal integer superspin supercurrent
of a chiral supermultiplet \cite{Buchbinder:2018nkp,Buchbinder:2018wzq}.
\item For the Poincar\'e case, we find that the supertrace vanishes and there is a family of consistent supercurrents given by \eqref{JP}.
Similar to the conformal case, the supercurrent exist for all values of $\Ysf$ but now there is no selection rule for $s$. It holds for all 
values of $s$. Also, one can set $\Ysf=0$ and recover the result for Poincar\'e integer superspin supercurrent of chiral supermultiplet
as described in \cite{Buchbinder:2018wzq}.
\end{enumerate}
Moreover, by projecting \eqref{csc}, \eqref{JP} and their superspace conservation equations to components, one can extract the corresponding 
higher spin currents  and their spacetime conservation equations. This procedure has been demonstrated in \cite{Buchbinder:2017nuc,
Koutrolikos:2017qkx,Buchbinder:2018wzq}. These component currents depend on 
$\W_{\a(2Y)}\Bigr|_{\substack{\th=0\\ \thd=0}}$ and $\D_{\b}\W_{\a(2Y)}\Bigr|_{\substack{\th=0\\ \thd=0}}$, which are the
bosonic and fermionic higher spin field strengths. The obtained higher spin currents will fall in the family of higher spin currents
presented in \cite{Berends:1985xx,Gelfond:2006be}.

In a previous work \cite{Buchbinder:2018wzq}, similar types of interactions were studied for the half-integer superspin supermultiplet
$[{(s+\ff1-2)}-\Ysf-\Ysf]$.
It is interesting to notice that consistent interactions for that case have in a sense an ``opposite'' behavior to what we find for the integer 
superspin case. For the half-integer superspin case
the value of $s$ is arbitrary, whereas the superspin $\Ysf$ had an upper bound $\Ysf\leq\ff{s}-2$.

Recently \cite{Alexander:2019vtb}, the effect of supersymmetric higher spin particles in
cosmological observables, such as the non-Gaussianity of the cosmic microwave background
was studied. The contribution of higher spin supermultiplets to the curvature perturbation
3-point function, originates from a wide class of interactions which allows the exchange of
higher spin particles and their superparteners. A subset of these interactions may include
the higher spin supercurrents presented in this work and \cite{Kuzenko:2017ujh,Buchbinder:2017nuc,
Hutomo:2017phh,Hutomo:2017nce,Koutrolikos:2017qkx,Buchbinder:2018wwg,
Buchbinder:2018nkp,Hutomo:2018tjh,Buchbinder:2018gle,Buchbinder:2018wzq}.

{\bf Acknowledgments}\\[.1in] \indent
The research of S.\ J.\ G.\ and K.\ K.\ is supported by the 
endowment of the Ford Foundation Professorship of Physics at 
Brown University. Also this work was partially supported by the U.S. National 
Science Foundation grant PHY-1315155.

\end{document}